\documentclass[epsfig,12pt]{article}
\usepackage{epsfig}
\usepackage{graphicx}
\usepackage{array}
\usepackage{caption}
\usepackage{subcaption}
\usepackage{slashed}
\usepackage{amsmath}
\usepackage{braket}

\newcommand{\beq}{\begin{equation}}   
\newcommand{\eeq}{\end{equation}}
\newcommand{\beqn}{\begin{eqnarray}}   
\newcommand{\eeqn}{\end{eqnarray}}

\newcommand{\bea}{\begin{eqnarray}}
\newcommand{\eea}{\end{eqnarray}}
\newcommand{\be}{\begin{equation}}
\newcommand{\ee}{\end{equation}}
\newcommand{\bead}{\begin{aligned}}
\newcommand{\eead}{\end{aligned}}

\newcommand{\nn}{\nonumber}
\newcommand{\pd}{\partial}

\def\ntwo{${\mathcal N}=2\;$}

\def\none{${\mathcal N}=1\;$}

\newcommand{\gsim}{\lower.7ex\hbox{$
\;\stackrel{\textstyle>}{\sim}\;$}}
\newcommand{\lsim}{\lower.7ex\hbox{$
\;\stackrel{\textstyle<}{\sim}\;$}}
\begin{document}

\begin{titlepage}

\begin{flushright}
FTPI-MINN-14/40, UMN-TH-3411/14\\
\end{flushright}

\vspace{0.4cm}

\begin{center}
{  \large \bf  Confining Strings in Supersymmetric Theories \\[2mm]
with Higgs Branches}
\end{center}
\vspace{0.3cm}

\begin{center}
 {\large 
 M. Shifman,$^a$ Gianni Tallarita,$^b$ and Alexei Yung$^{a,c}$}
\end {center}

\vspace{0.1mm}
 
\begin{center}

$^a${\em William I. Fine Theoretical Physics Institute, University of Minnesota,
Minneapolis, MN 55455, USA}\\[1mm]
$^b${\em Centro de Estudios Cient\'{i}ficos (CECs), Casilla 1469, Valdivia, Chile}
\\[1mm]
$^{c}${\em Petersburg Nuclear Physics Institute, Gatchina, St. Petersburg
188300, Russia}

\end {center}

\vspace{0.5cm}

\begin{center}
{\large\bf Abstract}
\end{center}  

We study flux tubes (strings) on the Higgs branches in supersymmetric 
gauge theories.  In generic vacua on the Higgs branches strings were shown to  develop  long-range ``tails'' 
associated with massless fields, a characteristic feature of the Higgs branch (the only exception is 
the vacuum at the base of the Higgs branch).
A natural infrared regularization for the above tails is provided
by a finite string length $L$. 

We perform a numerical study of these strings in generic vacua.
We focus on the simplest example of 
strings in \none supersym\-metric QED with the Fayet-Iliopoulos term.  
In particular, we examine the accuracy of a logarithmic approximation 
(proposed earlier by Evlampiev and Yung) for the tension of 
such string solutions. In the Evlampiev-Yung formula
the dependence of tension on the string length is logarithmic and the 
dependence on the geodesic length from the 
base of the Higgs branch is quadratic. We observe a remarkable agreement of our  
numerical results for the string tension with
the Evlampiev-Yung analytic expression.

\hspace{0.3cm}

\end{titlepage}


\section{Introduction}

Supersymmetric gauge theories provide an excellent theoretical laboratory for understanding strongly
coupled non-Abelian dynamics. In particular, the  dual
Meissner effect as a mechanism of confinement suggested in the mid-1970s \cite{mandelstam} was first analytically observed in 1994  in the framework of 
\ntwo supersymmetric theories
\cite{SW1,SW2}. The main feature of this mechanism is formation of the Abrikosov-Nielsen-Olesen
(ANO) \cite{ANO} flux tubes (confining strings). If in a given vacuum  quarks  condense then the
conventional magnetic ANO strings are formed. They confine monopoles. If, instead, monopoles condense,
the electric ANO strings are formed. They confine quarks \cite{SW1,SW2} (see also 
\cite{SYrev,SYdualrev} for reviews of scenarios with confined monopoles). 

Quite often supersymmetric gauge theories have Higgs branches. These are flat directions of the 
scalar potential on which charged scalar fields can develop vacuum expectation values (VEVs) breaking the gauge symmetry.
In many instances this breaking  provides  topological reasons behind  formation of the ANO strings.
The dynamical side of the problem of the confining string formation  in the theories 
with Higgs branches was addressed
in \cite{Penin:1996si,Yung:1999du,Evlampiev:2003ji}.  {\em A priori} it is not clear at all 
whether or not  stable string solutions 
exist  in this class of theories. The point is that the theories with a
Higgs branch represent a limiting case of type I  superconductor, with vanishing
Higgs mass. In particular, it was shown in \cite{Penin:1996si} that infinitely long strings cannot be formed 
in this case due to infrared divergences. 

Later this problem was studied \cite{Yung:1999du} in a more realistic  confinement setup, namely, 
 the string in question was assumed to have a large but finite length $L$. Finite length provides an infrared
regularization implying  \cite{Yung:1999du} 
that finite-size ANO strings still exist on the Higgs branches. They
become logarithmically ``thick'' due to the presence of massless fields and give
a confining potential for two heavy trial charges of the form
\beq
V(L) \sim \frac{L}{\log{L}}.
\label{confpot1}
\eeq
Note that $V(L) $, instead of being linear in 
separation $L$ is modified by $\log ( L\Lambda )$ in the denominator. 

The potential between heavy trial charges provides us with an order parameter  marking
distinct phases with different dynamical behaviors. Thus,  we see that theories with the Higgs branches
develop a novel confining phase with logarithmically nonlinear potential (\ref{confpot1}).

 Formation of strings in more generic theories with non-flat Higgs branches
 curved by the presence of the Fayet-Iliopoulos (FI) term was considered later
in \cite{Evlampiev:2003ji}.
In this case string profile functions can be approximated by an almost BPS core built from 
massive fields and a long-range ``tail'' built from massless fields. In this approximation the confining potential 
for the  simplest case of the $U(1)$ \none supersymmetric  gauge theory with one flavor of 
charged matter was shown to be
\beq
V(L) \sim L\left(1+ \frac{l^2}{\log{L}}\right),
\label{confpot2}
\eeq
where $l$ is the length of the geodesic line on the Higgs branch between 
the given vacuum and the base point of the Higgs branch.

In this paper 
our task is to  confirm the onset of the regime (\ref{confpot2}) for sufficiently large $L$.
This will allow us to better understand the limits of applicability of the
analytic consideration in \cite{Evlampiev:2003ji}.
To this end 
we numerically  study the string solution in 
 \none supersymmetric QED with the FI term. We find the string profile functions and
calculate the string tension. 
In agreement with the analytic formula (\ref{confpot2}) we observe that 
our numeric solution reproduces (with good accuracy) both features:  the logarithmic dependence
of the ``tail'' tension  on $L$ and the quadratic dependence on $l$.

The paper is organized as follows.  In Sec. \ref{sec2} we briefly review  a basic 
construction of length-$L$ flux tubes on curved Higgs branches in $\mathcal{N}=1$ SQED. 
Then we summarize main results concerning the  analytic approximation \cite{Evlampiev:2003ji} for their tension in terms of 
the distance from the base of the Higgs branch and $L$. In  Sec. \ref{sec3}
we obtain the  full numerical 
result for the profile functions of the string solution 
following  the general guidelines of \cite{Yung:1999du}. We then put the analytical tension formula (\ref{confpot2}) to  
test. Our numeric data establishes the
onset of the analytic approximation (\ref{confpot2}). In Sec. \ref{sec4} we present some conclusions.

 \label{intro}
\section{Flux tubes on curved Higgs branches}
\label{sec2}

\subsection{\none supersymmetric QED}

We begin by reviewing the  construction of flux tubes on curved Higgs branches in the 
Abelian gauge theories  \cite{Yung:1999du,Evlampiev:2003ji}.  
The starting point is $\mathcal{N}=1$ SQED with the action
\be
S_{\rm QED} = \int d^4x \left(\frac{1}{4g^2}F^2_{\mu\nu}+|D_\mu q|^2+
|D_\mu \tilde{q}|^2+V(q,\tilde{q})\right)
\label{qed}
\ee
where the covariant derivative is defined as
\be
D_\mu = \partial_\mu-\frac{i}{2}A_\mu
\ee
and the complex scalar fields $q$ and $\tilde{q}$ have opposite charges under the $U(1)$ gauge symmetry. We assume the charges for the scalar fields $n_e$ to be $|n_e|= 1/2$. The scalar potential is
\be
V(q,\tilde{q})=\frac{g^2}{8}\left(|q|^2-|\tilde{q}|^2-\xi\right)^2\,.
\label{pot}
\ee
It is obtained from the Fayet-Iliopoulus (FI) coupling for the $U(1)$ vector superfield with FI parameter $\xi$ after its auxiliary field $D$ is integrated out.

This model has a rich vacuum structure dictated by the vacuum condition
\be\label{condition}
|\braket{q}|^2-|\braket{\tilde{q}}|^2-\xi=0,
\ee
which describes a Higgs branch of dimension two: two complex scalars subject to one 
constraint after reduction  of a gauge phase. As is clear from the condition (\ref{condition}) in the vacuum the scalar fields develop vacuum expectation values thus completely breaking the $U(1)$ gauge symmetry. Correspondingly the photon acquires the mass
\be
m_\gamma = \frac{1}{2}g^2 v^2, 
\label{photonmass}
\ee
where
\be
v^2 = |\braket{q}|^2+|\braket{\tilde{q}}|^2.
\label{vev}
\ee
The scalar mass matrix has three zero eigenvalues corresponding to one ``eaten" combination and two massless 
scalar components of chiral multiplets living on the Higgs branch. In addition,  the mass matrix has one non-zero eigenvalue 
corresponding to a massive scalar field which is the superpartner of the massive vector 
supermultiplet, with mass equal to the mass of the photon $m_H = m_\gamma$.

Consider now the low-energy effective action for the theory (\ref{qed}), see  \cite{Evlampiev:2003ji}.
To   integrate out all massive fields in (\ref{qed}), namely, the  photon and the heavy scalar,
we use the following parametrization of the Higgs branch:
\bea
q &=& \sqrt{\xi}\,e^{i(\alpha +\beta)}\,\cosh(\rho)\,,\\[2mm]
\bar{\tilde{q}} &=& \sqrt{\xi}\,e^{i(\alpha-\beta)}\, \sinh(\rho)\,,
\label{higgsbranch}
\eea
where $\rho(x)$, $\alpha(x)$ and $\beta(x)$ are three real fields parametrizing $q$ and $\tilde{q}$
subject  to condition  (\ref{condition}). Once the gauge field is massive at low energies we can neglect its
kinetic term and eliminate  $A_{\mu}$ using the algebraic equation
\beq
A_{\mu} = -i\frac{\bar{q}\pd_{\mu} q -\pd_{\mu}\bar{q} q+ \tilde{q}\pd_{\mu} \bar{\tilde{q}} 
- \pd_{\mu}\tilde{q} \bar{\tilde{q}}}{|q|^2 +|\tilde{q}|^2} = 2\left(\pd_{\mu}\alpha 
+ \frac{\pd_{\mu}\beta}{\cosh{2\rho}}\right).
\eeq
Substituting this into the action (\ref{qed}) we arrive at
\beq
S_{\rm eff} = \xi \int d^4 x \, \cosh{2\rho}\,\left\{ (\pd_{\mu}\rho)^2 + (\pd_{\mu}\beta)^2\,\tanh^2{2\rho}
\right\}.
\label{leaction}
\eeq
This is the low energy-action in SQED, see (\ref{qed}),   containing only massless fields 
on the Higgs branch. The gauge phase $\alpha(x)$ is canceled out as expected.

In the simplest case, at the base of the Higgs branch, the vacuum is 
\be
\braket{\tilde{q}} =0, \quad \braket{q} = \sqrt{\xi}\,.
\label{base}
\ee
Far away from the base  we can parametrize vacua on the Higgs branch  as follows:
\bea
\label{limit}
\braket{q} &=& \sqrt{\xi}\,e^{i\beta_0}\,\cosh(\rho_0)\,,
\nonumber\\[2mm]
\braket{\bar{\tilde{q}}} &=& \sqrt{\xi}\,e^{-i\beta_0}\, \sinh(\rho_0)\,.
\eea
Here $\rho_0 = \rho(\infty)$ is a real dimensionless parameter describing how far the given vacuum  
lies from the base 
of the Higgs branch at $\rho_0 =0$, while $\beta_0$ is the residual phase which cannot be gauged away.
Each vacuum on the Higgs branch is characterized by two parameters $\rho_0$ and $\beta_0$.

\subsection{String solutions}

Consider first the vacuum (\ref{base}) located on the base of the Higgs branch.
This vacuum admits the standard Abrikosov-Nielsen-Olesen  vortices of infinite length \cite{ANO} in 
which the phase of the scalar field $q$ winds while its absolute value  rapidly tends to its 
vacuum expectation value at spatial 
infinity. These strings are BPS saturated, with the tension 
\be
T_{\rm BPS} = 2\pi n \xi\,.
\ee
Here $n$ the winding number of the solution. Below we consider elementary strings with $n=1$.

As was mentioned, in this paper we are interested in the flux tube solutions at a generic point on the Higgs branch. 
Such solutions can be found through the  procedure of dividing the radial separation from the string center into
two distinct spatial domains suggested in \cite{Evlampiev:2003ji}.  
 
First, one can safely assume that the photon field and the massive scalar field will form a BPS core of 
a finite radius determined by their common mass, namely, $$R_{\rm c}\sim 1/g\sqrt{\xi}\,.$$
This implies that for $r \leq R_{\rm c}$ we can look for the
solutions in which  $\tilde{q}\approx 0$. This domain is described by the standard BPS ANO string for 
which $T= T_{\rm BPS}$. 

Second, outside the above core, at  $r \geq R_c$,  the photon field 
vanishes. However, the
massless fields are excited, and their dynamics is determined by the low-energy action (\ref{leaction}). 
This leads to a long-range logarithmic tail,  contributing  both, to the profile functions and the string  tension 
 \cite{Evlampiev:2003ji}. 
 
The above long-range logarithmic 
tails require an infrared (IR) regularization. This statement is equivalent to the well-known 
result that the infinite-length strings are not allowed on Higgs branches \cite{Penin:1996si}.  
We will regularize\,
our solutions by considering strings of a finite length $L$.

The finite length IR regularization is physically motivated because it
corresponds to considering the string in the confinement setup. Namely,
we assume that finite length string is stretched between infinitely heavy
trial monopole and antimonopole at separation $L$. As we already mentioned the problem with 
infinite string arises because at large $r$ outside the
string core scalar fields satisfy free equations of motion and therefore, have
logarithmic behavior in two dimensions. Now for the case of the finite length string scalar 
fields also have logarithmic tails for 
$R_c\ll r\ll L$. However, as $r$ becomes of order of the string length $L$
the problem becomes three-dimensional rather then two-dimensional,
see \cite{Yung:1999du} for details. In three dimensions solution of the 
free equation of motion for the scalar field behaves as $1/|x_n|$ 
(rather then $\log{r}$),
where $x_n$, $n=1,2,3$ are the coordinates in the three-dimensional space.
These solutions can reach their boundary values at infinity dictated by (\ref{limit}). Thus, $1/L$
 plays the role of the IR regularization for the logarithmic behavior of scalar fields at large $r$.
In other words the finite length $L$ along the string axis translates into the IR regularization in the plane orthogonal to the string axis.

The total tension of the finite-$L$ solutions will  be given by
\be\label{ttot}
T = T_{\rm BPS}+ T_{\rm tail}\,,
\ee
where $T_{\rm tail}$ denotes the contribution to the tension from the long-range tail. It is given by
\beq
 T_{\rm tail} = \xi \int d^2 x \, \cosh{2\rho}\,\left[\rule{0mm}{4mm}
  (\pd_{i}\rho)^2 + (\pd_{i}\beta)^2\,\tanh^2{2\rho}
\right],
\label{tail}
\eeq
where we assume that the string  is a static solution aligned along the $x^3$ axis,
so the string profile functions in (\ref{tail}) depend only on coordinates $x^i$ with $i=1,2$,
if $r\ll L$.

Although the  tail  profile function were not found in  \cite{Evlampiev:2003ji} it was 
shown that the tail tension is determined by the universal formula depending on 
 the 
length $l$ of the geodesic line from the given vacuum to the base of the Higgs branch. In our model this 
length reduces to
\be
l = \int_0^{\rho_0} \sqrt{\cosh(2\rho)}\, d\rho\, ,
\ee
where the upper limit is the position of the vacuum on the Higgs branch, see (\ref{limit}).
The final result for strings of length $L$  (in the limit $L \gg R_{\rm c}$) is
\be
T_{\rm tail} \approx \frac{2\pi  \xi}{\log\left(g\sqrt{\xi}L\right)}\,l^2\,,
\ee
 see \cite{Evlampiev:2003ji} for a detailed derivation.
Hence, the expression for the total tension (\ref{ttot}) is   
\be\label{tension}
\frac{T}{2\pi\xi} \approx 1+\frac{1}{\log\left(g\sqrt{\xi}L\right)}l^2\,.
\ee
Formation of such strings leads to confinement of monopoles with the confining potential 
(\ref{confpot2}). It is not strictly linear in $L$. 

Another IR regularization more suitable for numerical calculations is to lift the Higgs branch
giving massless fields a small mass without breaking \none supersymmetry. One particular way to do this
is considered in \cite{Evlampiev:2003ji}. One can start from \ntwo QED and deform it with the mass term 
$\mu$ for the  neutral chiral multiplet. This term breaks \ntwo supersymmetry down to \none and at large
masses $\mu$ the deformed theory flows to \none QED. Integrating out the massive neutral multiplet
one obtains the scalar potential
\be
V(q,\tilde{q})=\frac{g^2}{8}\left(|q|^2-|\tilde{q}|^2-\xi\right)^2\, + \frac{1}{4\mu^2}
(|q|^2 +|\tilde{q}|^2)\,\left|q\tilde{q}-\frac{\eta}{2}\right|^2,
\label{potreg}
\ee
where $\eta$ is a new parameter which we take to be real, see \cite{Evlampiev:2003ji,Vainshtein:2000hu}
for detailes. We will consider this 
potential as an IR regularization of the one in (\ref{pot}). The Higgs branch is now lifted and we have 
an isolated vacuum with the vacuum value $\rho_0$ given by
\beq
\sinh{2\rho_0}=\frac{\eta}{\xi}
\label{rho0}
\eeq

The light scalar fields $\rho$ and $\beta$ in the low-energy action (\ref{leaction}) are no longer massless.
They acquire the mass
\beq
m_L= \frac{v^2}{2\mu},
\label{mL}
\eeq
where $v$ is the VEV given by (\ref{vev}). In terms of parameters of the potential (\ref{potreg})
$v$ can be expressed as 
\beq
v^4=\xi^2 + \eta^2.
\eeq

The relation between the two IR regularizations introduced above is 
\beq
m_L \sim \frac{1}{L}\, ,
\label{mLL}
\eeq
and the result (\ref{tension}) for the string tension reads
\beq
\label{ten}
\frac{T}{2\pi\xi} \approx 1+\frac{l^2}{\log\left(g\sqrt{\xi}/m_L\right)}\,.
\eeq
We use the latter IR regularization for the numerical calculations below.
This regularization allows us to consider infinitely long string and look for
solutions for the string profile functions in $(x^1,x^2)$ plane.

\section{Numerical solutions}
\label{sec3}

In this section we will construct full numerical solutions describing strings at a 
generic point on the Higgs branch and, with these solutions in hand, we can  directly verify the validity of 
the Evlampiev-Yung analytic formula (\ref{ten}). Our numerical solver involves a second order central finite 
difference procedure with accuracy $\mathcal{O}(10^{-4})$. From here on we set $$g=1\,.$$
It is convenient to define dimensionless quantities as
\be
\rho = \sqrt{\xi}r\,, \quad \tilde{\mu}^2 = \frac{\mu^2}{\xi}, \quad \tilde{\eta} = \frac{\eta}{\xi}.
\ee
Then the energy minimization equations, after using the {\em ansatz}
\bea
A_0 &=& A_r =0\,, \qquad A_\theta = 2(1-f(\rho))\,,
\nonumber\\[1mm]
q &=& \sqrt{\xi}q(\rho)e^{i\theta}\,,
\nonumber\\[1mm]
\tilde{q}&=&\sqrt{\xi}\tilde{q}(\rho)e^{-i\theta}\,,
\eea
reduce to
\bea\label{eom1}
&& q''+\frac{q'}{\rho} =\frac{1}{\rho^2}q f^2+\frac{1}{4}\left(q^2-\tilde{q}^2-1\right)q
\nn\\[3mm]
&&
+\frac{1}{4\tilde{\mu}^2}\left(q\tilde{q}-\frac{\tilde{\eta}}{2}\right)\left[q\left(q\tilde{q}-
\frac{\tilde{\eta}}{2}\right)+\tilde{q}\left(q^2+\tilde{q}^2\right)\right]\,,
\nn \\[3mm]
 \label{eom2}
&& \tilde{q}''+\frac{\tilde{q}'}{\rho} =\frac{1}{\rho^2}\tilde{q} f^2
-\frac{1}{4}\left(q^2-\tilde{q}^2-1\right)\tilde{q}
\nn \\[3mm]
&&
+\frac{1}{4\tilde{\mu}^2}\left(q\tilde{q}-\frac{\tilde{\eta}}{2}\right)\left[\tilde{q}\left(q\tilde{q}-\frac{\tilde{\eta}}{2}\right)+q\left(q^2+\tilde{q}^2\right)\right]\,,
\nn\\[3mm]
\label{eom3}
&& f''=\frac{1}{2}f\left(q^2+\tilde{q}^2\right)+\frac{f'}{\rho}\,  ,
\eea
where prime denotes differentiation with respect to $\rho$ and $\theta$ is the polar angle in $(x^1,x^2)$
plane.  

For large regularization parameter $\tilde{\mu}$, far from the base of the Higgs branch, where $f=0$, the solution  is basically determined by the Higgs constraint
\be
q^2-\tilde{q}^2-1=0\,.
\ee
Then, as is easily seen from Eqs. (\ref{eom1}), the fields $q$ and $\tilde q$ obey the
free equations of motion,
\be
(\rho q\,')'=0,\quad (\rho\tilde{q}\,')'=0\,,
\ee
with the standard logarithmic solutions. Correspondingly, the tension of the flux tube will be dominated by this large logarithmic tail.  Numerically the strategy 
is the following: the IR regularization is implemented as a mass regularization on the 
scalar fields, as explained in section 2.  Then, once we 
fix $m_L$ (making sure that $m_L << m_\gamma $) we impose boundary conditions on the fields at a 
fixed radial distance $R >> 1/ m_L$.  In this 
scheme, in which we fix $R$ we must ensure that $\rho_0$ is sufficiently small so that the BPS 
core approximation holds. If $\rho_0$ becomes too large then the $\tilde{q}$ field will develop in 
the core and spoil the theoretical approximation.

We are interested in solutions of (\ref{eom1}) with the following boundary conditions:
\bea
q(0)&=&\tilde{q}(0)=0\,,
\nonumber\\[1mm]
q(R) &=& \cosh(\rho_0), \qquad \tilde{q}(R)=\sinh(\rho_0)\,,
\nonumber\\[1mm]
f(0)&=& 1, \quad f(R)=0\,.
\eea

Figure \ref{2aaa} includes reference tables for the numerical values of the parameters $l$ and $\log(\sqrt{\xi}/m_L)$ for characteristic values of $\rho_0$ and $m_L$ used below. Solutions for the field profiles are shown in Figures 2 and 3 for varying values of $m_L$ and $\rho_0$. We fix $\sqrt{\xi}R=120$. Some important expected features can be seen in these plots: there is a BPS core formed by the photon field and the field $q$;  in this domain the field $\tilde{q}$   almost vanishes;
outside the BPS core the gauge field vanishes, and the massless scalar fields exhibit large logarithmic tails. \begin{figure}
\centering
\begin{subfigure}{.4\textwidth}
  \centering
  \begin{tabular}{| l |c| c | r| }
    \hline
   $\rho_0$&  $l$  \\ \hline\hline
    0.2&0.20    \\ \hline
    0.4&0.42   \\ \hline 
     0.6& 0.67 \\ \hline 
      0.8& 0.96 \\ \hline 
       1.0& 1.32 \\ \hline 
  \end{tabular}
  \label{fig4a}
  \caption{}
\end{subfigure}%
\begin{subfigure}{.6\textwidth}
  \centering
   \begin{tabular}{| l | c | r| r|r| }
    \hline
   $m_L/\sqrt{\xi}$ & $\log(\sqrt{\xi}/m_L)$ & $\tilde{\mu}_{0.2}$ & $\tilde{\mu}_{0.3}$ \\ \hline\hline
       0.05 & 2.99 & 10.81&11.85 \\ \hline
     0.033 &3.40 & 16.22&17.78\\ \hline 
     0.025 &3.69 &21.62&23.71\\ \hline 
     0.020 &3.91 &27.03&29.64\\ \hline
      0.017 &4.09 &32.43&35.56\\ \hline
  \end{tabular}
  \label{figtable}
  \caption{}
\end{subfigure}
\caption{\small Numerical values of (a) $l$ and (b) $\log(\sqrt{\xi}/m_L)$ for characteristic parameters used in the numerical solutions. We put $g = 1$. $\tilde{\mu}_{0.2}$ and $\tilde{\mu}_{0.3}$ show the values of $\tilde{\mu}$ at $\rho_0 = 0.2$ and $0.3$ for the values of $m_L$ used in the table.}
\label{2aaa}
\end{figure}
\begin{figure}[h]
\centering
\begin{subfigure}{.9\textwidth}
\centering
  \includegraphics[width=0.9\linewidth]{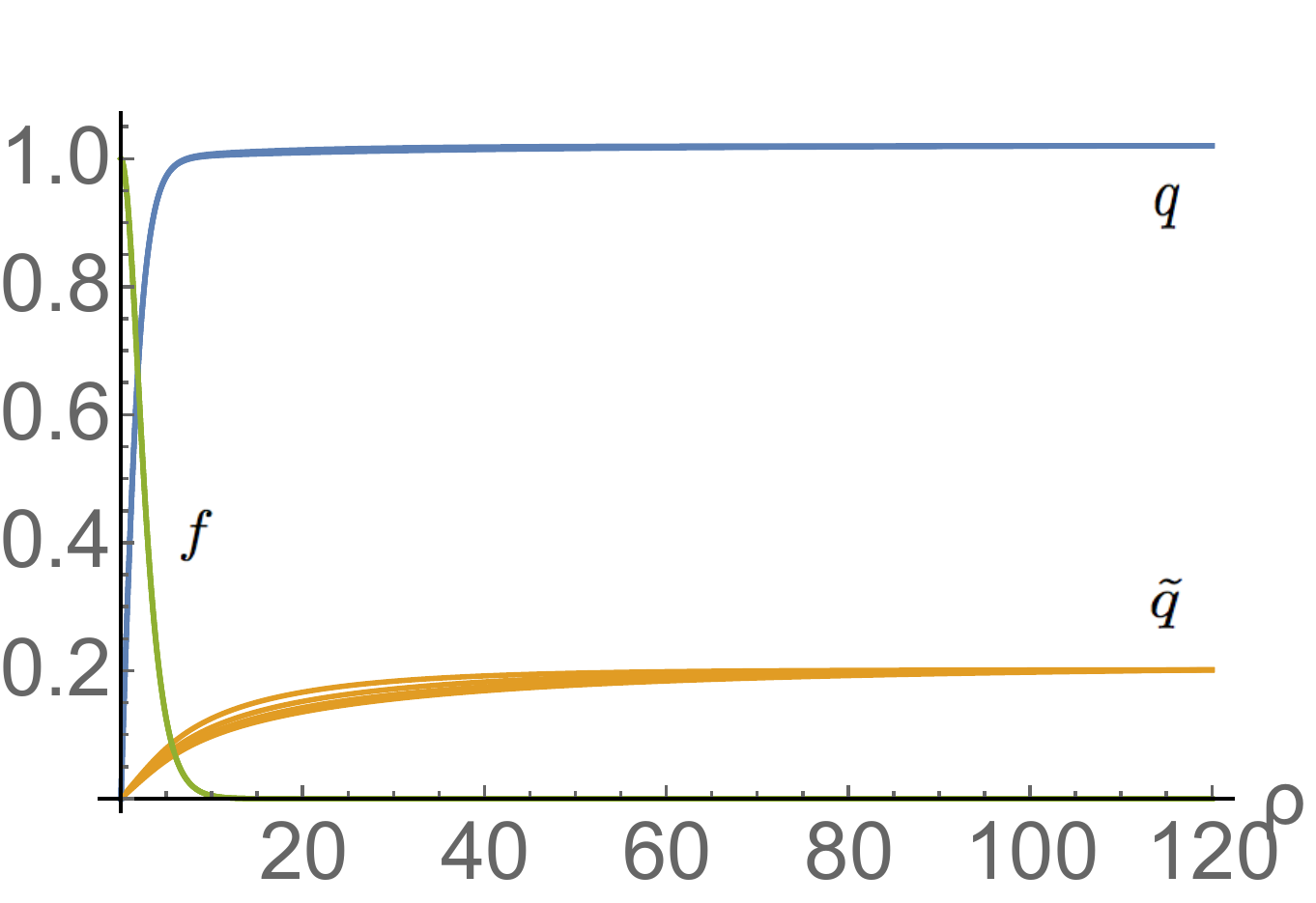}
  \caption{}
  \label{fig1}
\end{subfigure}%
\\
\begin{subfigure}{0.9\textwidth}
\centering
  \includegraphics[width=0.9\linewidth]{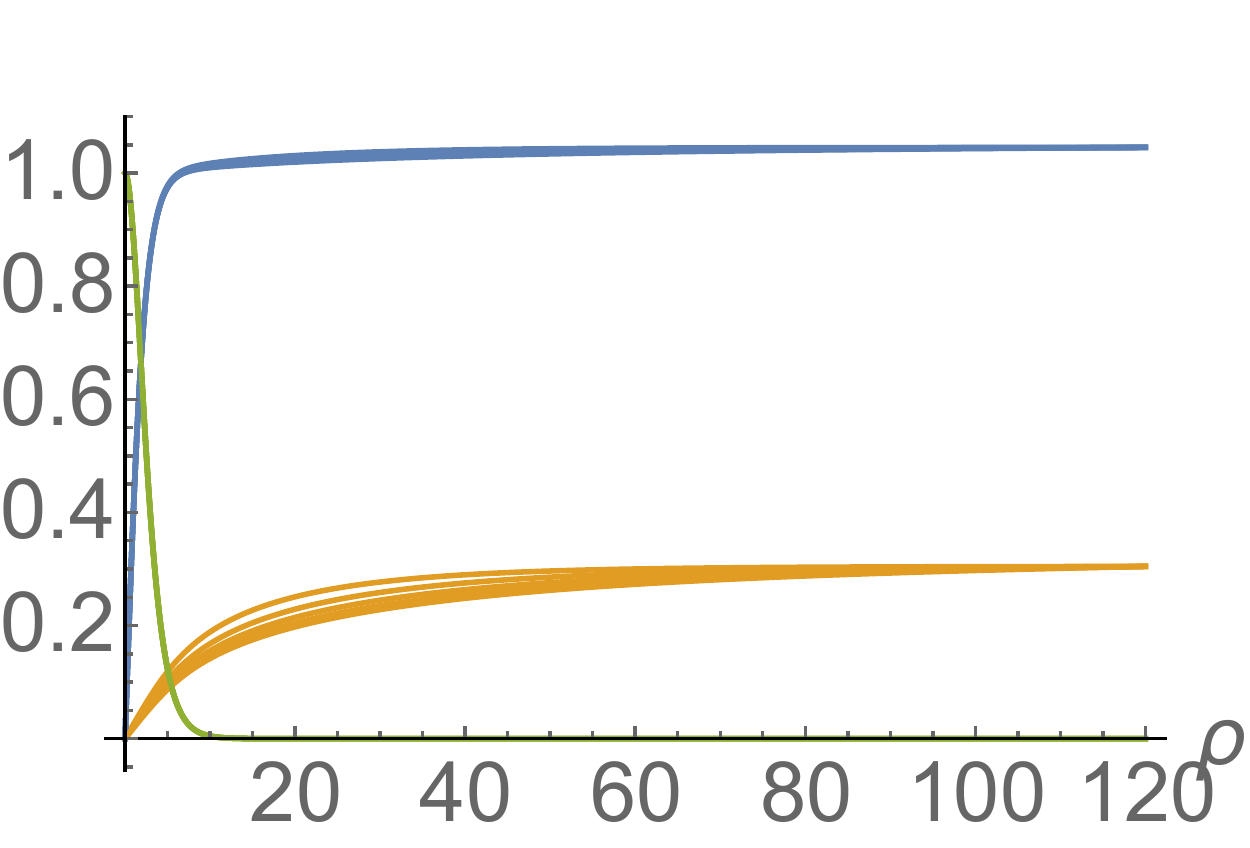}
  \caption{}
  \label{fig2}
  \end{subfigure}
  \caption{\small Numerical solutions for field profile functions varying $m_L$, the curve labels in (a) also apply to plot (b). In (a) we use $\rho_0=0.2$ and in (b) we pick $\rho_0=0.3$. The values of $m_L$ used are reported in Figure 1.}
\end{figure}

\begin{figure}
\centering
  \includegraphics[width=0.9\linewidth]{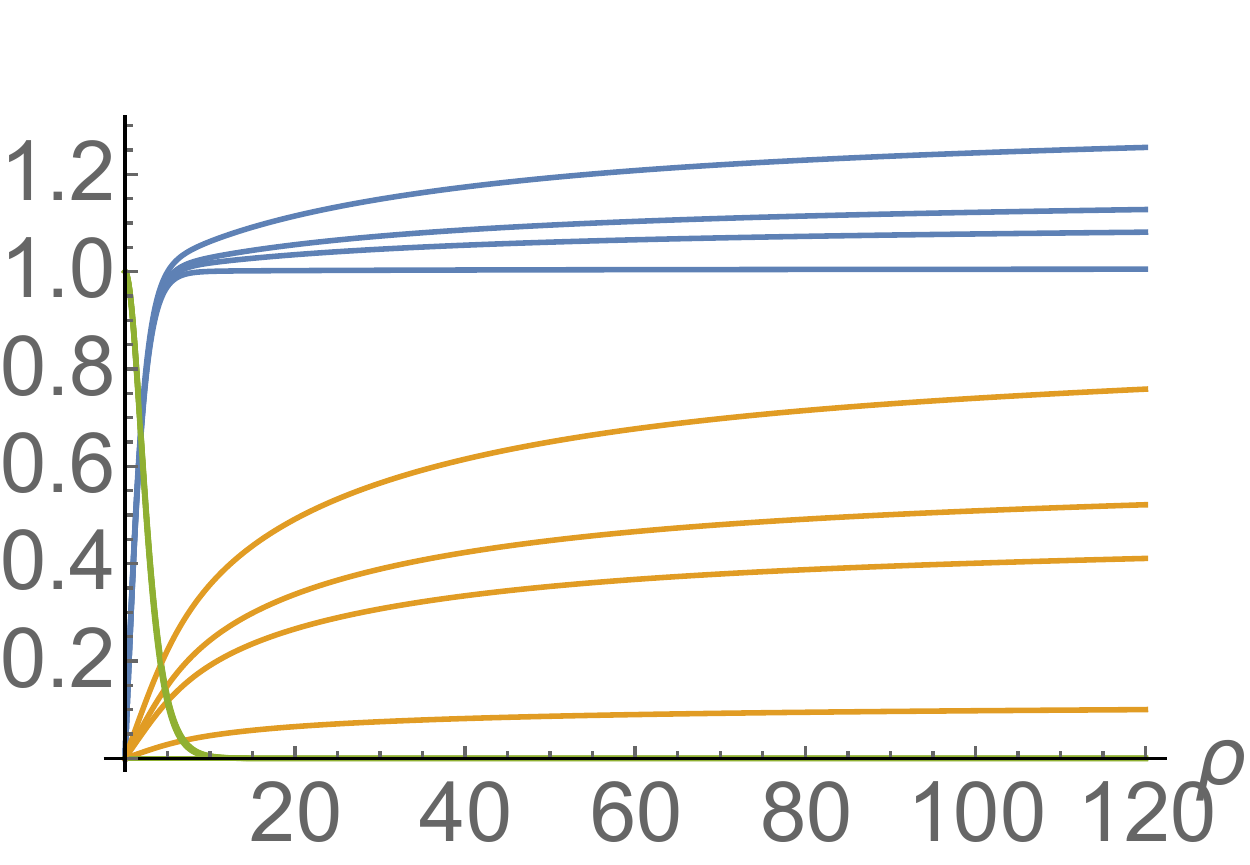}

  \caption{\small Field profiles varying $\rho_0$ at $m_L =  0.017$. The plots correspond to $\rho_0 = 0.1, 0.3, 0.4, 0.7$.}
  \label{figgg3}
\end{figure}

\begin{figure}
\centering
  \includegraphics[width=\linewidth]{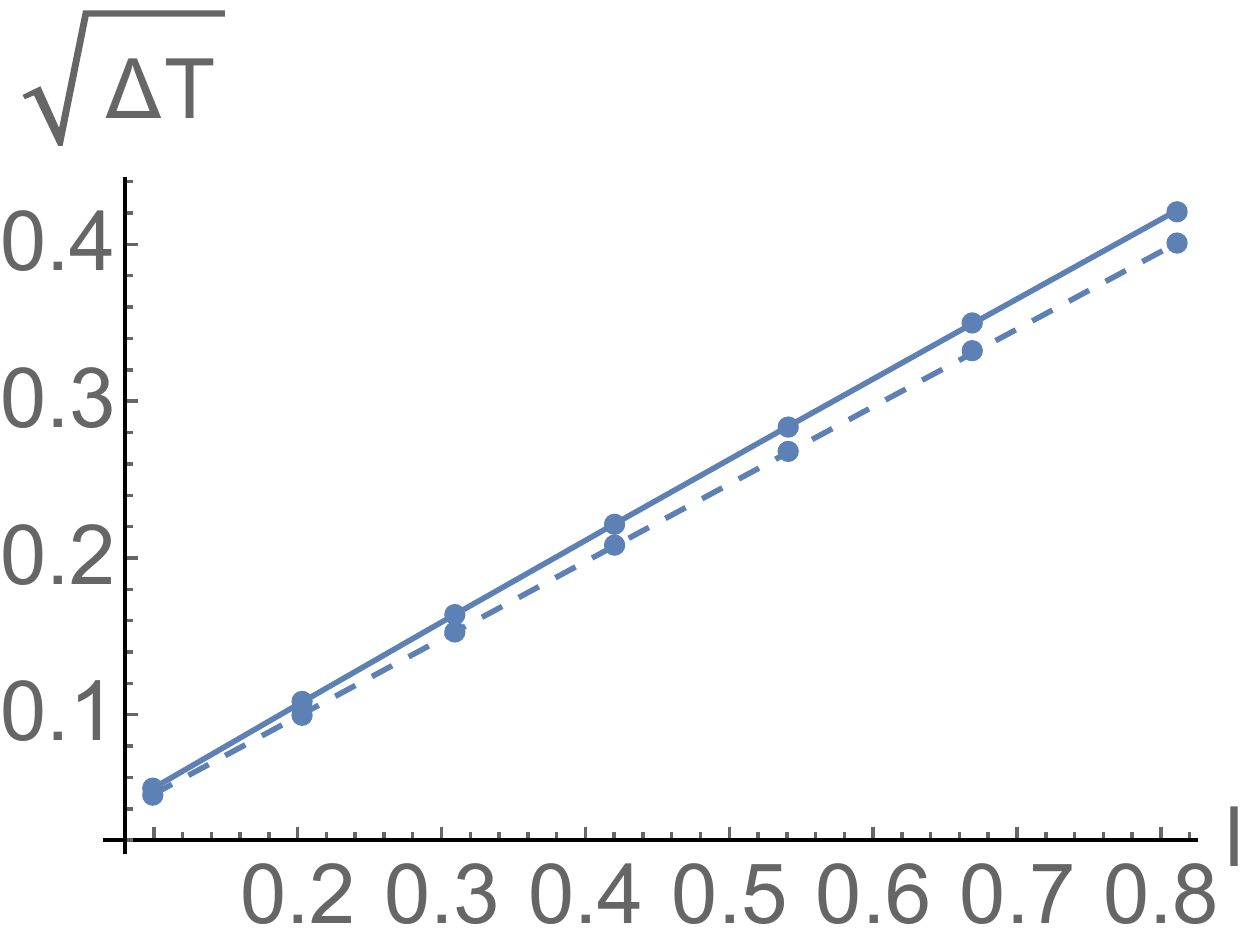}
 \caption{\small Difference between numerical and theoretical tensions $\Delta T/\xi$ for varying $l$. Solid line corresponds to numerical result, dashed line to theoretical. The values of $\rho_0$ used are 0.1 to 0.7 in steps of 0.1.}
 \label{figgg4}
\end{figure}

\begin{figure}
\centering
\begin{subfigure}{0.9\textwidth}
\centering
  \includegraphics[width=\linewidth]{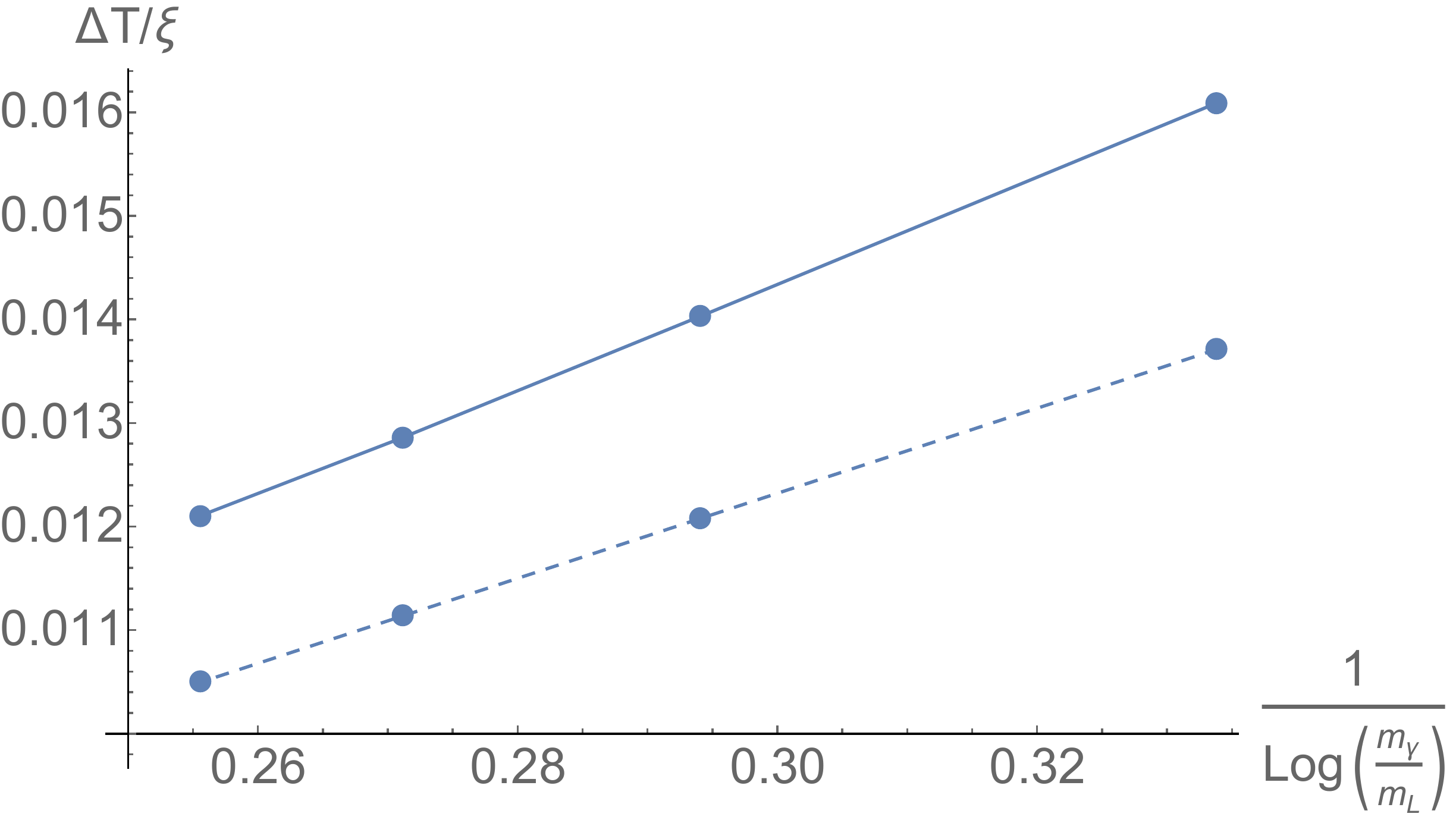}
  \caption{}
  \label{fig6}
\end{subfigure}%
\\
\begin{subfigure}{0.9\textwidth}
\centering
  \includegraphics[width=\linewidth]{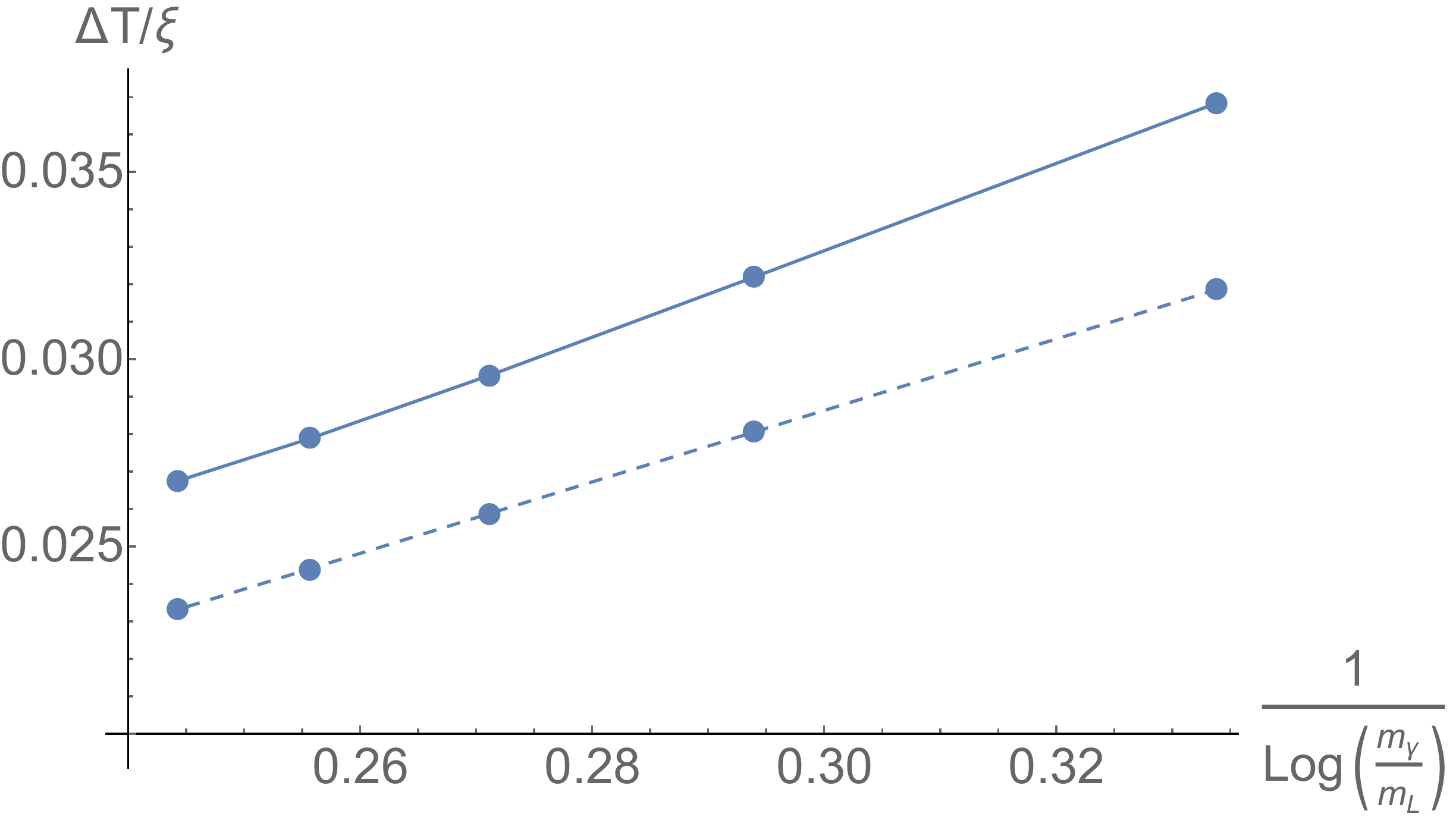}
  \caption{}
  \label{last}
\end{subfigure}%
\caption{\small Difference between numerical and theoretical tensions $\Delta T$ for varying $m_L$. Solid line corresponds to numerical result, dashed line to theoretical. (a)   $\rho_0 = 0.2$.  (b) $\rho_0 = 0.3$. The values of $m_L$ used are reported in Figure 1, plot (a) excludes the point in which our numerical procedure could not determine a solution.}
 \label{figgg5}
\end{figure}

Figures \ref{figgg4} and \ref{figgg5} show the results of the numerical analysis of the tension formula (\ref{tension}). As  seen from the plots, at larger values of the parameters we find that for some particular combinations the accuracy of our procedure is not enough to find a solution. These points are excluded from the plots. The results involve the difference between the numerical result for the tension and its BPS part coming from the core,
\be
\frac{\Delta T}{\xi} = \frac{T - T_{\rm BPS} }{2\pi\xi}= \frac{1}{\log\left(g\sqrt{\xi}/m_L\right)}\,l^2\,  .
\ee
In particular, Fig. \ref{figgg4} shows a plot of $\sqrt{\Delta T/\xi}$ in which we fix $m_L$ and vary $l$. We observe 
a number of 
 important features. 
 
 First, the numerical and theoretical results coincide (within numerical accuracy) at $\rho_0 =0$.  This is expected,
 of course, since at this point we are at the base of the Higgs branch and the tension coincides with the BPS result. As we move along the Higgs branch by increasing $\rho_0$ we see an increasing disagreement between the numerical solution and the analytic (approximate) theoretical prediction. Once again, this is expected as in this domain one picks up large $l$ effects. 
 Second, we observe that the numerical solution for $\sqrt{\Delta T/\xi}$ is a linear function of $l$, in perfect agreement with the theoretical expression. A slight deviation in the slope can be explained by the  logarithmic accuracy
in the denominator in the theoretical prediction, the ${\log\left(\sqrt{\xi}/m_L\right)}$ term in the denominator
can be shifted by a constant non logarithmic term of the order of unity. In fact, this conjecture is supported by the subsequent plots.
 
Figure \ref{figgg5} shows similar plots in which we fix $\rho_0$ in order to verify the logarithmic dependence on $m_L$. Once again, by observing the linear dependence in the plot, we verify that the logarithm dependence up to small values of $m_L$. Given that the theoretical approximation is for large values of logarithm
in (\ref{ten}) it 
is not surprising to find a better agreement as $m_L$ decreases (and thus $1/\log(\sqrt{\xi}/m_L)$ decreases). Quantitatively we find that at the smallest value of $m_L$ the agreement between the theoretical $T_{\rm their}$ and numerical $T_{\rm number}$ results, respectively is 
\be
\frac{T_{\rm numer} - T_{\rm theor}}{T_{\rm numer}} \times 100 \approx 10 \%,
\ee
(this value holds for both values of $\rho_0$ investigated).
The agreement is quite satisfactory given the magnitude of the parameters used at this point (see Fig. \ref{2aaa}). 
Adding a non-logarithmic term in the denominator of the theoretical expression and fitting it
we could have dramatically improved the agreement (by two orders of magnitude!).

Indeed, since Eq. (\ref{tension}) is an approximation with logarithmic accuracy we propose a simple modification of this formula which we can test numerically. Let us replace (\ref{tension}) by
\be
\frac{T }{2\pi\xi} \approx 1+\frac{1}{\log\left(\sqrt{\xi}/m_L\right)-c}\, l^2\, ,
\label{31}
\ee
where $c$ is a constant to be fitted numerically. We find that, for $\rho_0 = 0.2$ and $R=120$ 
\be
\frac{T_{\rm numer} - T_{\rm theor}}{T^c_{\rm numer}} \times 100 \approx 0.4\%\,,
\ee
provided that $$c\approx 0.55\,.$$
In other words, the value of the non-logarithmic constant in (\ref{31}) turns out to be less than one, a complete success. For values of $\rho_0$ greater than those used in the plots we find that one cannot ignore the effects of the $\tilde{q}$ field in the core.

\section{Conclusions}
\label{sec4}

In this paper we analyzed  magnetic flux tubes (strings) on the Higgs branch  in supersymmetric 
QED.  In generic vacua on the Higgs branches these strings were previously shown to  develop  long-range  tails 
due to massless fields existing on the Higgs branch. A natural infrared regularization for the above tails can be provided by a finite string length $L$. Numerically a small supersymmetry preserving mass regularization was used, the two being related by $m_L \sim 1/L$.

We performed a detailed numerical analysis of  flux tube solutions at generic points on the Higgs
 branch in $\mathcal{N}=1$ SQED. We found numerical solutions for the field profile functions defining 
such strings that (i) contain a BPS core and (ii) besides the core contain long logarithmic tails due to 
the massless scalar fields characteristic to  the Higgs branch. Using these solutions we then analyzed the 
Evlampiev-Yung analytic formula  (\ref{ten})  (presenting the small-$m_L$ geodesic approximation for 
the tail contribution to the string tension)  comparing it to numerical results .
We found good agreement for the predicted functional dependence on $m_L$  and $l$.

 \section*{Acknowledgments}

This work  is supported in part by DOE grant DE-SC0011842 and Fondecyt grant No. 3140122. G.T. would like to thank the William I. Fine Theoretical Physics Institute  of the University of Minnesota for hospitality during  completion of this work.
The work of A.Y. was  supported by William I. Fine Theoretical Physics Institute  of the  University of Minnesota,
by Russian Foundation for Basic Research under Grant No. 13-02-00042a and by Russian State Grant for
Scientific Schools RSGSS-657512010.2. The work of A.Y. was supported by Russian Scientific Foundation 
under Grant No. 14-22-00281. The Centro de Estudios Cient\'{i}ficos (CECS) is funded by the Chilean
Government through the Centers of Excellence Base Financing Program of Conicyt.

\end{document}